\documentclass[aps,prd,amsmath,nofootinbib,groupedaddress,showkeys,showpacs]{revtex4-2}
\usepackage{graphicx}
\def\qt{\ln q^{\frac{1}{3}}}

\def\3G{^{(3)}\mathcal{G}}
\newcommand{\be}{\begin{equation}}
\newcommand{\en}{\end{equation}}
\newcommand{\eps}{{\rlap{\lower2ex\hbox{$\,\,\tilde{}$}}{\epsilon_{ijk}}}}
\newcommand*{\epss}{{\rlap{\lower2ex\hbox{$\,\tilde{}$}}{\epsilon}}\,}
\begin{document}
\title{Cosmic Time And The Initial State Of The Universe}
\author{Chopin Soo}\email{cpsoo@mail.ncku.edu.tw}
\affiliation{Department of Physics, National Cheng Kung University, Taiwan}
\date{\today}
\begin{abstract}
The exact solution of the Hamiltonian constraint in canonical gravity and the resultant reduction of Einstein’s theory reveal the synergy between gravitation and the intrinsic cosmic clock of our expanding universe.
Intrinsic Time Geometrodynamics advocates a paradigm shift from four covariances to just spatial diffeomorphism invariance.
Consequently, causal time-ordering and quantum Schr\"odinger–Heisenberg evolution in cosmic time become meaningful. The natural addition of a Cotton–York term to the physical Hamiltonian changes the initial data problem radically. In the classical context, this is studied with the Lichnerowicz–York equation; quantum mechanically, it lends weight to the origin of the universe as an exact Chern–Simons Hartle–Hawking state, which features Euclidean–Lorentzian instanton tunneling. At the level of expectation values, this quantum state yields a low-entropy hot smooth Robertson–Walker beginning in accord with Penrose’s Weyl Curvature Hypothesis.
The Chern–Simons Hartle–Hawking state also manifests transverse traceless quantum metric fluctuations, with, at the lowest approximation, scale-invariant two-point correlations as one of its defining characteristics.
\end{abstract}
\keywords{Canonical Quantum Gravity, Intrinsic Time Geometrodynamics, Penrose Weyl Curvature Hypothesis, Lichnerowicz-York equation, Gravitational Chern-Simons functional, Hartle-Hawking state\\
{\bf Published}: {\it Universe\,\,} {\bf 2023}, Volume 9, Issue 12, 489}
\maketitle
\section{The Cosmic Clock and Intrinsic Time Geometrodynamics}

The union of gravity and quantum mechanics is needed to address the physics of the very early universe and to elucidate the microscopic beginning of our universe in conditions of extreme temperatures and curvatures.
However, the road to quantum gravity is obstructed by formidable conceptual and technical challenges.
General Relativity (GR)  is a theory about spacetime,
 yet a straightforward notion of ``time'' does not come readily; it is even plagued with the ``problem of time'' \cite{Isham1,Isham2,Isham3}.
The standard Arnowitt–Deser–Misner (ADM) \cite{ADM} canonical formulation of Einstein's theory reveals a total Hamiltonian that is, apart from a possible boundary term, made up solely of constraints.
 There is no ready generator of  ``time'' translation; and no immediate variable to be identified as  ``time''.    ``The label $x^0$  itself is irrelevant, and `time' must be determined intrinsically''
 was\linebreak DeWitt's \cite{DeWitt} conclusion, and suggestion that a degree of freedom (d.o.f.) of the theory should be identified as ``time'' to deparameterize the dynamics.
 DeWitt's suggestion has an additional technical meaning. In his seminal work \cite{DeWitt}, the d.o.f. chosen as the time is $q^{\frac{1}{4}}$ where $q$ is the determinant of the spatial metric.
 (3+1)-decomposition of spacetime separates four-geometry into intrinsic entities pertaining to each spatial hypersurface and extrinsic quantities, which describe how a spatial hypersurface is curved in spacetime; therefore, DeWitt's choice is ``intrinsic''. In Intrinsic Time Geometrodynamics (ITG) \cite{SOOYU,SOOYU1,ITGBook} and this present article, ``intrinsic time'' is advocated in the same context of arising from the spatial metric.

 Canonical quantization of GR based on the ADM formulation leads to the Wheeler-DeWitt equation \cite{DeWitt,Wheeler}; however, the intrinsic time variable in the equation is spatially dependent, thus multi-fingered. This poses a challenge to the mutual consistency of the infinitude of possible time orderings.
 The equation turns the Hamiltonian constraint of GR into a dynamical quantum equation for three-geometry (equivalence class of spatial metric modulo spatial diffeomorphisms).
 The absence of four-covariance in the Wheeler–DeWitt formulation of quantum geometrodynamics is clear; Wheeler also emphasized that Einstein's geometrodynamics is the dynamics of three-geometry, {\it not four-geometry}~\cite{Wheeler}.
Even earlier, an unequivocal call to abandon four-covariance in GR was sounded by Dirac, one of the founding fathers of quantum mechanics, in Ref. \cite{Dirac2}.
The simplicity of the Hamiltonian analysis and the fact that only the spatial metric, rather than the full spacetime metric, is dynamical led Dirac to conclude and declare that ``four-dimensional symmetry is not a fundamental symmetry of the physical world.''
Regarded by many as an exemplar of succinctness in his writings, Dirac nevertheless expressed the same sentiment in no less than five instances in the paper: once in the abstract and four times in the conclusion. Two of these instances were even highlighted in italics!
Further clues that indicate four-covariance is neither needed nor even present in classical GR can be found in the initial data formulation of York \cite{extrinsic}. Therein, the Hamiltonian constraint is treated as an equation to uniquely fix the conformal factor when the trace of the extrinsic curvature is made spatially constant and adopted as the global time variable (this is the formalism of extrinsic time)\footnote{Appendix \ref{AA} contains related discussions; and a comparison of York's extrinsic time formulation and ITG can be found in Ref.\cite{DOF}.}.
Instead of being regarded as a generator of ``gauge symmetry'',  the Hamiltonian constraint is thus used to eliminate a d.o.f.  from the theory.  ITG parallels this but with a global time variable that is intrinsic and a resultant physical Hamiltonian that generates global cosmic time translation.
In ITG, the Hamiltonian constraint is solved explicitly by eliminating the trace of the momentum in terms of the other variables \cite{SOOYU,SOOYU1,ITGBook}.

A key obstacle to the viability of Einstein's theory and its four-covariant extensions as renormalizable perturbative quantum field theories was pointed out by Horava\cite{Horava}.
This lies in the deep conflict between unitarity and spacetime general covariance. Renormalizability of GR can be improved and achieved through the introduction of higher derivative terms, but space–time covariance also requires higher time derivatives, as well as spatial derivatives, of the same order, thus compromising the stability and unitarity of the theory.
Horava's bold proposal is to keep unitarity at the expense of relinquishing four-covariance, to retain only spatial diffeomorphism symmetry at the fundamental level.
The proposed modification to Einstein's theory is to introduce higher spatial, but not time, derivatives of the metric to improve ultraviolet convergence.
Without the complication of higher time derivatives, unitarity can be protected, with GR recovered at low spatial curvatures and low frequencies.
In a theory with only spatial diffeomorphism invariance, the preeminent term to add \cite{Horava} to the Hamiltonian involves the trace of the square of Cotton–York tensor \cite{York-Cotton1,York-Cotton2,extrinsic}.
This tensor is transverse and traceless and contains up to three spatial derivatives of the metric (further details can be found in Appendix \ref{AB}). It is special to three dimensions, and it would never have been introduced had four-covariance been the overriding concern.

ITG actualizes \cite{ITGBook}, in a new vista, Dirac's conviction that four-covariance is not fundamental to GR, York's parallel of using the Hamiltonian constraint to both eliminate a d.o.f. from the theory and obtain a reduced physical Hamiltonian, and the consistent incorporation of Horava's proposal of modifying Einstein's theory with higher spatial curvature terms. It recognizes the primacy of a Cotton–York term in overcoming the many technical challenges of quantum gravity.
Some of the ramifications this brings, including its impact on the origin of the universe, are studied in this work.

In Einstein's geometrodynamics, the spatial metric and its conjugate momentum, $(q_{ij}, {\tilde \pi}^{ij})$, are the fundamental canonical variables.
This bequeaths the theory with several advantages, including a well-defined concept of a spacelike hypersurface---in Wheeler's words, ``a notion of `simultaneity' and a common moment of a rudimentary `time''' \cite{Wheeler}, an elemental ingredient that is needed in imposing the quantum microcausality relations of spacelike local commutativity or anticommutativity of fields.
A closed universe \cite{Silk1,Silk2,Silk3} comes with a finite spatial volume $V$; an ever-expanding universe yields the direction of cosmic time.
In ITG, $d\ln V$ plays the role of an intrinsic time interval; time is inherent in the framework rather than external.
The expanding closed universe itself is the all-encompassing and infinitely robust clock that renders ``time'' in gravitation concrete, universal, and comprehensible, while its own dynamics, origin, and destiny are dictated by the laws of gravitation and quantum mechanics.

Instead of a theory obfuscated by four-covariant paradigm, Schr\"odinger–Heisenberg quantum evolution and causal temporal ordering with respect to the cosmic clock of our expanding closed universe is well defined in a theory of gravitation with just spatial diffeomorphism invariance.
The ITG paradigm shift prompts modifications of Einstein's theory to include a Cotton–York term to improve ultraviolet convergence \cite{Horava,ITGBook}, in addition to the infrared convergence gained from spatial compactness.
The physical contents of Einstein's General Relativity are captured at low frequencies and low curvatures; except in the very early Cotton–York dominated era when the universe was tiny in size, the time-dependent physical Hamiltonian ensures the primacy of Einstein's theory and its observational consequences at later times of large spatial volumes.  This very time-dependence permits the universe to begin very differently from how it will end.

ITG also advocates an exact Chern–Simons Hartle–Hawking state as the quantum origin of the universe. At the level of expectation values, this is in accord with Penrose's Weyl Curvature Hypothesis,
of a smooth Robertson–Walker, but also hot (due to Euclidean instanton tunneling) beginning; the gravitational arrow of time of increasing intrinsic volume concurs with the thermodynamic Second Law arrow of increasing entropy.
A signature of the Chern–Simons Hartle–Hawking state is that it manifests, at the lowest-order approximation, scale-invariant two-point correlation function for transverse traceless quantum metric fluctuations.

\subsection{Symplectic Potential of Geometrodynamics, and Generalized DeWitt Supermetric}\label{section2.3}


 The symplectic potential of geometrodynamics has a remarkable decomposition,
\begin{equation}\label{symplecticpot}
 \int\, {\tilde\pi}^{ij}\delta q_{ij} = \int {\bar\pi}^{ij}\delta{\bar q}_{ij} + \tilde\pi \delta \qt;
\end{equation}
wherein the spatial metric $q_{ij} =q^{\frac{1}{3}}{\bar q}_{ij}$ is expressed in terms of its determinant, $q$, and unimodular part, ${\bar q}_{ij}$; and the momentum is split into its trace, $\tilde\pi := q_{ij}{\tilde\pi}^{ij}$, and traceless part, $\bar{\pi}^{ij}:= q^{\frac{1}{3}}   [\tilde{\pi}^{ij}  - \frac{q^{ij}}{3}{\tilde\pi}].$ The only nontrivial Poisson brackets are
 \begin{eqnarray}\label{piij}
    \{\bar{q}_{kl}(x),\bar{\pi}^{ij}(x')\}_{P.B.}&=&P^{ij}_{kl}\,\delta^3(x,x'), \\
     \{\bar{\pi}^{ij}(x),\bar{\pi}^{kl}(x')\}_{P.B.}&=& \frac{1}{3}({\bar q}^{kl}\bar{\pi}^{ij} - {\bar q}^{ij}\bar{\pi}^{kl})\delta^3(x,x'),
 \end{eqnarray}
and\footnote{The inverse of ${\bar q}_{ij}$ is denoted by ${\bar q}^{ij}$; also $\int\, \tilde\pi \delta \qt =\int\, \frac{1}{3}\frac{\tilde\pi }{q}\delta{q}$.}
\be\label{lnqpiPB}
\{{q}(x), {\tilde\pi(x')}\}_{P.B.} = {3}q \delta^3(x,x');
\en
with the trace-free projector, $P^{ij}_{kl} :=  \frac{1}{2}(\delta^i_k\delta^j_l + \delta^i_l\delta^j_k) - \frac{1}{3}\bar{q}^{ij}\bar{q}_{kl}$. So $(\tilde\pi, \qt)$ commutes with the barred variables $({\bar q}_{ij},{\bar\pi}^{ij})$; this permits a degree of freedom (d.o.f.), namely $\ln q$, separate from the others, to be identified as the carrier of temporal information and, as shall be elucidated, $\tilde\pi$ to play the role of the corresponding Hamiltonian density.

The generic ultralocal DeWitt supermetric \cite{DeWitt} compatible with three-covariance is
\be\label{supermetric}
G_{ijkl} = \frac{1}{2}(q_{ik}q_{jl} +  q_{il}q_{jk}) - \frac{\lambda}{3\lambda -1} q_{ij}q_{kl},
\en
with $\lambda$ as deformation parameter. An intrinsic clock is, in fact, not tied to four-covariance ($\lambda = 1$ in Einstein's GR) since the supermetric has signature $({\rm sgn}[\frac{1}{3}-\lambda], +,+, + ,+ ,+)$, which comes equipped as $(-, +,+,+,+,+)$ as long as $\lambda > \frac{1}{3}$. The single negative eigenvalue for $\lambda > \frac{1}{3}$ corresponds to the $\delta\ln q$ mode in $G^{ijkl}\delta q_{ij}\delta q_{kl}$ (with $G^{ijkl}$ being the inverse of $G_{ijkl}$).

 \subsection{Generalized Hamiltonian Constraint of Geometrodynamics; and Global Intrinsic Time Interval}\label{solveHC}

A generalized Hamiltonian constraint of geometrodynamics, which is quadratic in momenta, is
\begin{eqnarray}\label{GeneralH1}
0=H&=&  \frac{2\kappa}{\sqrt{q}}[G_{ijkl}\tilde{\pi}^{ij}\tilde{\pi}^{kl} + {\mathcal V}(q_{ij})]\nonumber\\
\label{FactH}&=&-\frac{2\kappa}{\sqrt{q}}[(\beta \tilde\pi- \bar{H})(\beta \tilde\pi + \bar{H})];
\end{eqnarray}
with
\begin{equation}\label{Hbar}
\bar {H}(\bar{\pi}^{ij}, \bar {q}_{ij}, q)
= \sqrt{ \frac{1}{2}[\bar{q}_{ik}\bar{q}_{jl} +\bar{q}_{il}\bar{q}_{jk}]\bar{\pi}^{ij}\bar{\pi}^{kl} + {\mathcal V}({q}_{ij})}\,;
\end{equation} and $\beta := \sqrt{\frac{1}{3(3\lambda - 1)}}$. The final step of \eqref{GeneralH1} is a noteworthy factorization.  In this context, Einstein's theory which corresponds  to $\lambda = 1$ or $\beta =\sqrt{\frac{1}{6}}$, and
 \be\label{Einsteinpotential}
 {\mathcal V}({\bar q}_{ij}, q) =- \frac{q}{(2\kappa)^2}[R - 2\Lambda ],
 \en
 is a particular realization of a wider class of theories \cite{ITGBook,SOOYU,DOF}.
 There is the remaining momentum constraint, $H_i = -2q_{ij}\nabla_k{\tilde\pi}^{jk}=0$, which generates spatial diffeomorphisms (or spatial Lie derivative changes) of the variables.

Although $q$ is a tensor density under spatial diffeomorphisms, the change $\delta\ln q =(\delta q)/q$ is scalar. For a closed three-manifold, Hodge decomposition of $\delta\qt$ yields
\be\label{YTT}
\delta\qt(x)   = \delta T + \nabla_i {\delta Y}^i(x),
\en
wherein $\delta T$  is harmonic, {\it independent} of $x$ and gauge-invariant under diffeomorphisms, whereas $\delta Y^i$ can be gauged away because, fortuitously, the Lie derivative $\mathcal{L}_{\delta N^i} \qt (x)= \frac{2}{3}\nabla_i\delta N^i(x)$.
Multiplying \eqref{YTT} with ${\sqrt q}$ and integrating over the spatial volume $V$, yields \cite{ITGBook} $dT= \frac{2}{3}d\ln V$  which is a {\it global} intrinsic time interval that is invariant under spatial diffeomorphisms.

\subsection{Cosmic Time and the Reduced Physical Hamiltonian}

Solving the Hamiltonian constraint explicitly by eliminating $\tilde\pi$ in terms of the rest of the variables and using constant-$T$ slicings reveal in ITG the synergy between dynamics of geometry and the cosmic clock of our expanding universe.
The solution of the constraint \eqref{GeneralH1} is ${\tilde\pi} =\pm \frac{1}{\beta}{\bar H}$. To wit, the action in ADM form \cite{ADM},
\begin{equation}
S=\int dT \int \left[{\bar\pi}^{ij}\frac{\partial{\bar q}_{ij}}{\partial T} +{\tilde\pi}\frac{\partial\ln q^{\frac{1}{3}}}{\partial T} - N^iH_i -NH \right]d^3x,
\end{equation}
 reduces (with ${\tilde\pi} =-\frac{1}{\beta}{\bar H}$)  to
 \begin{eqnarray}
 S_{reduced}&=&
\int dT\left[\int ({\bar\pi}^{ij}\frac{\partial{\bar q}_{ij}}{\partial T}  - N^iH_i)d^3x\right] -\int dT \left[\int (\frac{\bar H}{\beta}\frac{\partial\ln q^{\frac{1}{3}}}{\partial T})d^3x\right] \nonumber\\
&=&\int dT\left[\int {\bar\pi}^{ij}\frac{\partial{\bar q}_{ij}}{\partial T}d^3x  - \int (\frac{1}{\beta}{\bar H} + N^iH_i)d^3x\right].
\end{eqnarray}
In the above, with $\frac{\partial\ln q^{\frac{1}{3}}}{\partial T}=1$, and the reduced physical Hamiltonian which emerges in ITG is $H_{phys} =\frac{1}{\beta}\int{\bar H} d^3x$,
wherein ${\bar H}$ is as defined in \eqref{Hbar}.
In the reduction, the choice of ${\tilde\pi} =-\frac{1}{\beta}{\bar H}$ (instead of $+\frac{1}{\beta}{\bar H}$) is invoked to obtain a positive-definite Hamiltonian in an expanding, instead of contracting, universe.
Complementary derivations which lead to the same conclusions and reduced Hamiltonian have been elaborated on elsewhere \cite{ITGBook,SOOYU1,SOOYU,ITQG,CS}.

In place of the Dirac algebra \cite{Dirac} of Hamiltonian and momentum constraints, a physical reduced Hamiltonian corresponding to cosmic time $T$ is obtained.
Only the closed algebra of spatial diffeomorphism generated by $H_i$ remains.
The transition to the global Hamiltonian above naturally prompts the extension of Einstein's theory by modifying the potential without adding time derivatives. Horava's emphasis on the crucial role of the Cotton–York tensor \cite{Horava},
and other considerations discussed in detail in Ref. \cite{ITGBook}, lead the ITG framework to settle on
\begin{eqnarray}\label{sqrtformH}
H_{phys}&=&\frac{1}{\beta}\int{\bar H}d^3x \nonumber\\
&=&\frac{1}{\beta}\int \sqrt{  \bar{\pi}_{i}^{j}\bar{\pi}_{j}^{i} -\frac{q}{(2\kappa)^2}(R - 2\Lambda)+ g^2\hbar^2{\tilde C}^i_j {\tilde C}^j_i}\, \,d^3x.
\end{eqnarray}

In $H_{phys}$, Klauder's momentric variables \cite{Klauder}, ${\bar{\pi}}^{i}_{j}$, are intriguing entities, which, in lieu of utilizing the momenta, confer many advantages \cite{ITGBook}.
Classically, ${\bar{\pi}}^{i}_{j} :={\bar q}_{jk}{\bar{\pi}}^{ik}$. Quantum mechanically, it is advantageous to treat the momentric as fundamental rather than as a composite of the metric and momentum.
The momentric variables themselves generate an $sl(3,R)$ algebra at each spatial point; the kinetic operator in the Hamiltonian, $\bar{\pi}_{i}^{j}\bar{\pi}_{j}^{i}$, is a Casimir invariant; and on ${\bar q}_{ij}$ they generate $SL(3.R)$ transformations
\footnote{The explicit commutation relations are \cite{ITGBook}
$\bigl[{\bar q}_{ij}(x), {\bar \pi}^{k}_{l}(y)\bigr] =i\hbar (\frac{1}{2}\bigl(\delta^i_m\bar{q}_{jn}+\delta^i_n\bar{q}_{jm}\bigr)-\frac{1}{3}\delta^i_j\bar{q}_{mn})\delta^{3}(x,y)$;
$\bigl[ {\bar \pi}^{i}_{j}(x), {\bar \pi}^{k}_{l}(y)\bigr] = \frac{i\hbar}{2}\bigl(\delta^k_j{\bar \pi}^i_l-\delta^i_l{\bar \pi}^k_j\bigr)\delta^{3}(x,y)$. }.
 The momentric variable, $\bar{\pi}_{i}^{j}$,  is a density weight one tensor of dimension $\hbar/ L^3$.

 The Cotton–York tensor density, ${\tilde C}^i_j$, is of weight one and dimension $1/L^3$. It is the third order in spatial derivatives of the metric. $H_{phys}$ is the integral (over closed spatial manifold) of a density weight one spatial scalar; it is thus invariant under spatial diffeomorphisms (i.e., it commutes with the generator of spatial diffeomorphisms $H_i$). The Cotton–York tensor (density), ${\tilde C}^{ij} ={\tilde\epsilon}^{imn}\nabla_m (R^j\,_{n}-\frac{1}{4}\delta^{j}\,_nR)$, is special to three dimensions.
It is the functional derivative with respect to the spatial metric of the three-dimensional gravitational Chern–Simons invariant of the metric affine connection.
Its vanishing is the necessary and sufficient condition for conformal flatness. In four dimensions, this role is played by the Weyl tensor, which is identically zero in three dimensions.
A four-covariant extension of Einstein's theory incorporating the square of the Weyl tensor (and with dimensionless coupling in the extended action) is renormalizable; alas, loss of unitarity due to the presence of higher time derivatives comes with the Weyl extension.
With only spatial diffeomorphism invariance, the (trace of the) square of the Cotton–York tensor density has just the right scalar weight, the associated dimensionless coupling $g$, and higher spatial derivatives to be incorporated into the Hamiltonian of ITG.
${\tilde C}^i_j$ is conformal invariant; thus, it is a function of the unimodular spatial metric ${\bar q}_{ij} $, rather than of the full metric.
This independence from $q$ (hence the intrinsic time parameter) means that it dominates the early universe at small spatial volumes when the $qR$ and $q\Lambda$ terms in the Hamiltonian are negligible; whereas at late times and large volumes, the theory becomes Einsteinian. The Cotton–York tensor is precisely the gravitational counterpart of the Yang–Mills magnetic field.
Both are functional derivatives of their respective Chern–Simons invariant; the crucial difference is that the spatial metric is the fundamental variable in GR, whereas, in Yang–Mills, it is the gauge connection; consequently, the Cotton–York tensor is third order in spatial derivatives of the metric (see Appendix \ref{AB}). The incorporation of the Cotton–York tensor into the GR potential is as natural as the appearance of the magnetic potential in the renormalizable Yang–Mills theory.
These parallels are occluded in four-covariant paradigms of GR obsessed with Lagrangians and gravitational four-curvature invariants.

\section{Quantum Gravity as Schr\"odinger–Heisenberg Quantum Mechanics}

In place of the Wheeler–DeWitt equation \cite{DeWitt} (which is a constraint at each spatial point) and the difficulties it entails,
the Hamiltonian of ITG, $H_{phys} $ is a global entity that is integral over all spatial points; and cosmic volume time is global, not many-fingered.
The resultant quantum equation is the familiar first-order-in-time Schr\"odinger equation, which is an equation for cosmic time-development, not an equation at each spatial point; unitarity or conservation of norm holds provided the Hamiltonian of ITG is self-adjoint{\footnote{With suitable regularization, the Hamiltonian can be put \cite{ITGBook,ZPE} in the positive-definite self-adjoint form $\frac{1}{\beta}\int \sqrt{Q^\dagger Q + q{\cal K} }\,d^3x$.}.}
There is stupendous power and staggering simplicity in this equation. Due to its first-order dependence on time,
there is always a solution, no matter how complicated $H_{phys}$ may be (this includes time-dependent Hamiltonians), and the solution is unique given an initial state $\Psi(T_0)$.
The formal solution can be derived by integrating, yielding\linebreak $\Psi(T) = U(T,T_0)\Psi(T_0)$,  with
 \be
 U(T,T_0):={\cal T}(\exp\left[-\frac{i}{\hbar}\int^T_{T_0} {\hat H}_{phys}(T')dT'\right]).
 \en
 This is a $T$-ordered operator due to, in general, non-commutation of the Hamiltonian operator at different times.  A temporal ordering, in $T$,  of the state emerges.
 This time-development operator can be expanded explicitly as a time-ordered Dyson series \cite{ITGBook}. Global cosmic time interval and $H_{phys}$ are diffeomorphism invariant; the whole description is gauge-invariant if $\Psi(T_0)$ is also invariant under spatial diffeomorphisms (this is true of the explicit initial Chern–Simons Hartle–Hawking state which will be advocated).
 In ITG, the intrinsic time interval $dT$ is dimensionless, and  $H_{phys}$ is of dimension ${\hbar}$. What is paramount to causality is not the physical dimension of time but the existence of an ordering.
 Quantum gravity spells the demise of classical spacetime but not of temporal ordering.  The fundamental gauge-invariant entity that can be, and is, in fact, time-ordered, is the quantum state of the entire physical universe.
 Therein lies the substantiation of what remains of ``cause and effect''. No more, but no less, is allowed.

$H_{phys}$ contains no $\tilde\pi$, thus it commutes with $\ln q^{\frac{1}{3}}$; and $\frac{\partial \ln q^{\frac{1}{3}}}{\partial T}=1$ from the Hodge decomposition.
Consequently, the Heisenberg  equations of motion for operators yield
\be
\frac{d\ln q_\textsf{H}^{\frac{1}{3}}}{dT} =U^\dagger(T,T_0) (\frac{\partial \ln q^{\frac{1}{3}}}{\partial T} +\frac{1}{i\hbar} [\ln q^{\frac{1}{3}}, H_{phys}])U(T,T_0) =1;
\en
and $q_\textsf{H}(x,T) =q_0(x) e^{3(T-T_0)}$ is the solution. No ad hoc ``fixing'' of $q$ in $H_{phys}$ is necessary in the Heisenberg picture.
Time-dependence of the Hamiltonian comes from the $q$-dependence of  $q(R - 2\Lambda)$ in $H_{phys}$, while ${\tilde  C}^i_j $ is independent of $q$.  In terms of $q$ and ${\bar R}$ (the Ricci scalar curvature of ${\bar q}_{ij}$),
\be
R=q^{-1/3}[\bar{R}- 2{\bar q}^{ij}\bar{\nabla}_{i}\partial_{j}\ln q^{1/3}-\frac{1}{2}{\bar q}^{ij}(\partial_{i}\ln q^{1/3})(\partial_{j}\ln q^{1/3})];
\en
so the potential, $q(R - 2\Lambda)$ scales as $q^{2/3}$ for the $qR$ term and as $q$ for the cosmological constant term.  Hence, they scale exponentially with $T$, whereas the Cotton–York term is unaffected.

In the quantum context, if the physical Hamiltonian commutes an anti-unitary time reversal operator $\texttt{T}$, then whenever $\Psi(T)$ is a solution of the Schr\"odinger equation the time-reversed wave function, $\texttt{T}\Psi(T)$, satisfies $i\hbar\frac{d}{d(-T)}\texttt{T}\Psi(T)=H_{phys}\texttt{T}\Psi(T)$.
In Minkowski spacetime, time reversal is an improper Lorentz transformation; in a generic curved spacetime, it can be generalized to an orientation reversal which maps the vierbein one-form $e^0 \rightarrow -e^0$ while preserving the spatial dreibein one-forms $e^{a=1,2,3}$ (the four-metric $g_{\mu\nu}dx^\mu dx^\nu = -(e^0)^2 + e^a e^a$).   In quantum gravity, there is no recourse to four-dimensional classical spacetime; however, the Schr\"odinger–Heisenberg mechanics in ITG allows time reversal invariance or violation to be properly phrased in Wigner's formulation \cite{Wignertime1,Wignertime2} discussed above. In ITG, the unimodular spatial metric ${\bar q}_{ij}$ is invariant while the momentric is odd under time reversal (these assignments also preserve the fundamental commutation relations under $\texttt{T}$); $q$  on the other hand scales exponentially with intrinsic time $T$. The physical Hamiltonian, $H_{phys}(T)$, has explicit time-dependence\footnote{Time-dependence of a Hamiltonian is allowed in quantum mechanics; it does not spoil unitarity which is ensured by the self-adjointness of the Hamiltonian.} through $q$, and it is not an even function of the intrinsic time $T$. This means the Hamiltonian of ITG, $H_{phys}$ as in \eqref{sqrtformH}, violates time reversal invariance, and the theory is time-irreversible.

A time-dependent theory has an added feature: it can allow different physics to come into dominance at different times without having to postulate time-dependent or ever-changing fundamental laws.
An in-built time asymmetry, in theory, carries an arrow of time. In ITG, the Hamiltonian is Cotton–York dominated at early times of small spatial volumes and Einsteinian at late times and large volumes.
This agrees with the dominance and correctness of GR in the current epoch.  As the universe continues to expand, the cosmological constant term will become increasingly dominant.

\section{Origin of the Universe}

The incorporation in ITG of the Cotton–York tensor, which is unique to three dimensions, brings surprising new features that could not be foreseen within a four-covariant paradigm of GR.
The effect of a Cotton–York term in the extended Lichnerowicz–York equation in narrowing down the initial data of the universe will be addressed in detail in Appendix \ref{AA}.

\subsection{Bekenstein–Hawking Black Hole Entropy and Penrose’s Weyl Curvature Hypothesis}

Penrose has pointed out that there is indeed an asymmetry in our universe \cite{Nature of spacetime,Penrose1,Penrose2,Penrose1979, Penrose1989}.
Observations support the picture that our universe has a rather smooth Robertson–Walker-type early stage, and as it ages, structures from gravitational clumping turn into stars, galaxies, and black holes.
A Schwarzschild black hole is associated with Bekenstein–Hawking entropy \cite{Bekenstein1,Bekenstein2,Bekenstein3} $S_{bh}=\frac{1}{4}(\frac{A}{L^2_{Planck}})k_B$,
 wherein $A=4\pi r^2_s$ is the area of the Schwarzschild event horizon with $r_{s} =\frac{2GM}{c^2}$.
Compared to other objects, a black hole carries an inordinate amount of entropy. If an object of mass $M$ containing
$\frac{M}{m_B}$ number of baryons were to collapse into a black hole, the entropy per baryon would be $\frac{1}{4}[\frac{4\pi r^2_S}{L^2_{Planck}(M/m_B)}]k_B$.
This is roughly an entropy of $10^{20}k_B$ per baryon for $M =10M_\odot$. All other known forms of entropy are, by comparison, negligible.
If the more than $3\times10^{80}$ baryons in our universe were totally made up of such black holes, their entropy (in units of Boltzmann constant $k_B$) would exceed $10^{100}$, a googol!
Penrose estimates that the total entropy of a black hole-dominated universe could reach $S=10^{123} k_B$.
 Thermodynamic postulate of equal a priori probabilities for all the accessible microstates and the Boltzmann entropy relation $S= k_B\ln \Omega$  imply a staggering number of microstates $ \Omega  \sim e^{{10}^{123}}$, which is much greater than a googolplex ($10^{\rm googol}$), would correspond to black hole initial data. This carries the profound implication that black hole states would swamp all others, and extreme fine-tuning would be necessary for the universe to steer away from the preponderance of being born a black hole or from being black hole-dominated at the beginning. Thus, a smooth Robertson–Walker beginning without black holes would, in Penrose's words, be ``extraordinarily special''. He also emphasized that the argument does not need to rely on strict thermodynamic adherence. What can reasonably be argued on general physical grounds is a phase space that includes gravitational d.o.f. would quite certainly be predominantly black hole initial data.
 Compared to what it theoretically could have been in a black hole-dominated phase space, the entropy content of the cosmological gravitational field at the beginning was extremely low.
Robertson–Walker isotropic singularity is one in which the Ricci tensor dominates over the Weyl tensor, whereas black holes are associated with Weyl singularities.
Penrose links the initial low-entropy content of the universe with the effective vanishing of the Weyl curvature tensor near the Big Bang.
He proposed the Weyl Curvature Hypothesis of vanishing Weyl tensor as a selection rule for the initial data of our universe \cite{Nature of spacetime,Penrose1,Penrose2,Penrose1979,Penrose1989}.

In addition, the Hypothesis does imply from the initial {\it sans} black hole low-entropy condition to the formation of structure through clumping of matter into stars, black holes, galaxies, and gigantic black holes,
the overall entropy increases as the universe ages as black hole entropy by far outweighs all other forms of entropy\footnote{Even if all black holes were to eventually evaporate away after eons, it is possible a very small cosmological constant, ergo very large cosmological horizon, and entropy, remains. In a de Sitter manifold, the area of the cosmological horizon is inversely proportional to the value of the cosmological constant.}.
Such an initial condition would lead to an entropy arrow of time in accordance with the Second Law of Thermodynamics.
In ITG, in an ever-expanding universe, the cosmological arrow of time points in the direction of increasing volume. This gravitational–cosmological arrow of time and the thermodynamic arrow of time are aligned if the initial quantum state of the universe favors the Penrose selection rule.

An imposed condition on the Weyl tensor is essentially extraneous to Einstein’s field equations, which directly relate the Ricci tensor, rather than the Weyl curvature, to energy-momentum.
Although the Weyl Curvature Hypothesis proffers a tantalizing resolution of fundamental problems, the means to achieving and realizing it in a natural context within semiclassical and quantum Einstein theory remain elusive.
Penrose advocates the theory of Cyclic Conformal Cosmology wherein the final conformally flat (hence vanishing Weyl tensor)  stage of one universe is mapped onto the beginning of another universe together with the associated remnants of the past universe \cite{Penrose3,CCC1,CCC2}.
This is not without difficulties \cite{CCC-No1,CCC-No2,CCC-No3,CCC-No4}.
Furthermore, Penrose's Hypothesis is a classical statement that is prone to quantum and semiclassical corrections (as Hawking has pointed out \cite{Nature of spacetime}), especially at the beginning of the universe when quantum effects are expected to be significant for Planck size beginnings. In addition, the selection of initial data of vanishing Weyl curvature at the Big Bang is exceedingly unlikely within the classical context of a gravitational theory described entirely by Einstein's GR.
The strict vanishing of Weyl curvature is also problematic to the existence of primordial gravitational waves characterized by nontrivial Weyl invariants.
However, Cotton–York dominance brings new physical perspectives on the nature of the early universe and on the quantum origin of the universe.

\subsection{Hartle–Hawking No-Boundary Proposal}

As it is first order in time, Schr\"odinger–Heisenberg mechanics needs an additional physical input: the initial quantum state.
The nature of the Hamiltonian in the very early universe yields hints, but the initial condition of the universe demands further insights and input.

\begin{figure}[h]
\includegraphics[width=10.5 cm]{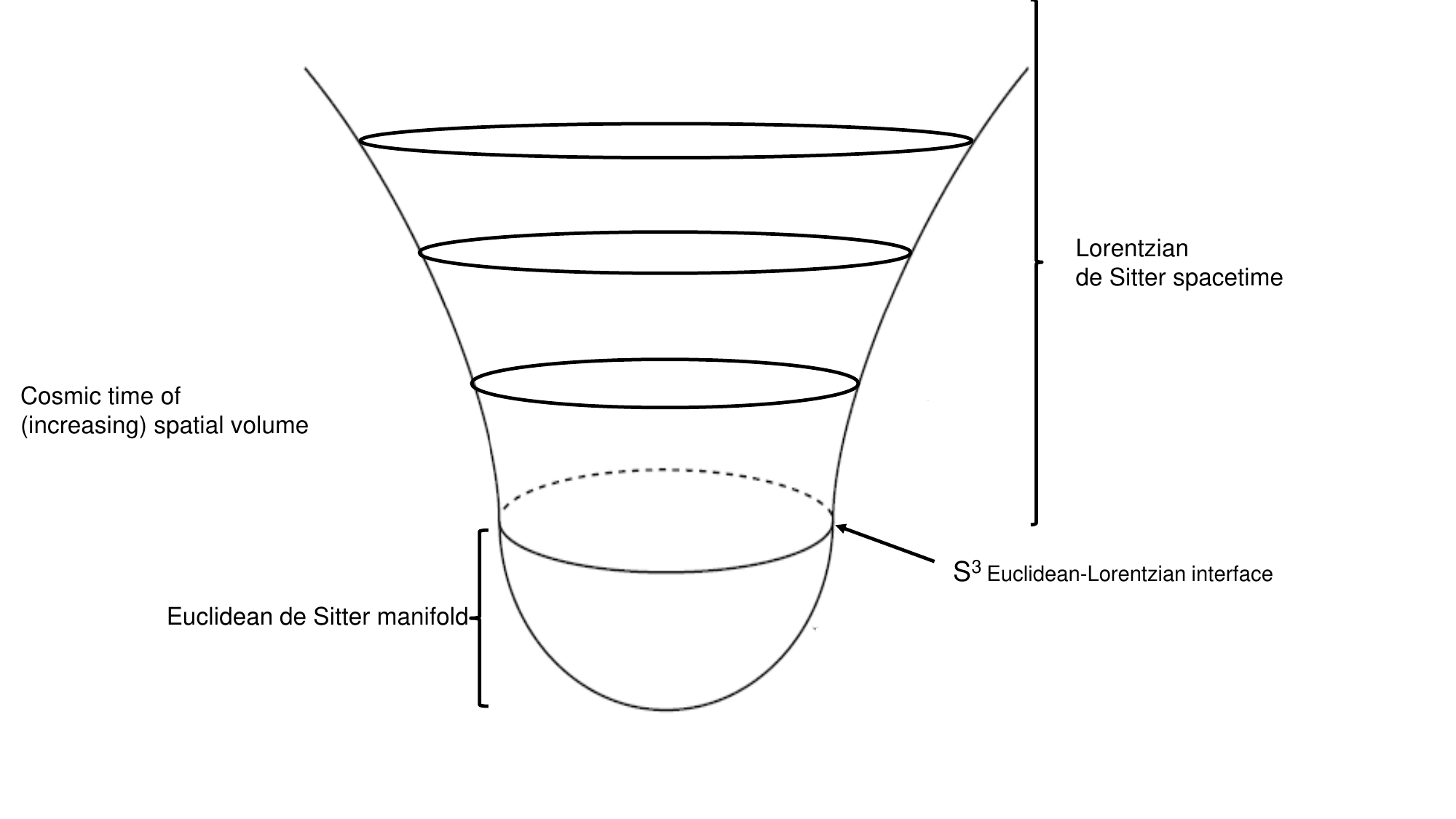}
\caption{Origin of the universe from Euclidean tunneling. (Semiclassical picture of the quantum Chern–Simons Hartle–Hawking state.)\label{fig1}}
\end{figure}

Hawking, in Ref. \cite{Nature of spacetime}, proposed a semiclassical picture of how our universe came into being.
It arose out of Euclidean tunneling from literally nothing, from a single point. To many, this is intuitive and appealing in its simplicity.
In Ref. \cite{Nature of spacetime}, there is an image of this Euclidean–Lorentzian tunneling for the de Sitter four-manifold of a Euclidean origin from a single point growing to the size of the Lorentzian de Sitter throat and then expanding exponentially onwards as a Lorentzian de Sitter spacetime (see Figure \ref{fig1}). This is an exact solution of Einstein's equations with a cosmological constant (due to the conformally flat $S^3$ spatial slicings, it is also a solution of ITG \cite{ITGBook}).\vspace{-6pt}

Hartle and Hawking advocated \cite{Hartle-Hawking} the no-boundary proposal (sometimes referred to, perhaps more appropriately, as the ``one-boundary proposal'') for the wave function and quantum origin of the universe. The proposal states that the wave function or probability amplitude on a closed spatial hypersurface $\Sigma$ is given by the exponential of the Euclidean gravitational action summed over all Euclidean manifolds with $\Sigma$ as their sole boundary. More precisely,
the Hartle–Hawking wave function for the origin of the universe is
\begin{equation}\label{HHInstanton}
\Psi[q^\Sigma_{ij}] =N\int Dg^{M}_{\mu\nu}\, e^{-S_E[g^{M}_{\mu\nu}]},
\end{equation}
wherein $S_E$ is the Euclidean action for gravity, and $\Sigma$, on which $q_{ij}$ is the induced spatial metric, is the only boundary ($\partial M=\Sigma$) of all the compact Euclidean four-geometries summed over.
The left-hand side is the wave function of the universe, so the functional integral over all compact four-geometries bounded by a given three-geometry is the amplitude for that three-geometry to arise from a zero three-geometry, i.e., a single point of zero volume.
``In other words, the ground state is the amplitude for the universe to appear from nothing'' \cite{Nature of spacetime}, and ``the boundary condition of the universe is that it has no boundary'' \cite{Hartle-Hawking}.

\section{Chern–Simons Initial State}

ITG takes the Hartle–Hawking proposal seriously and finds it amazingly in concord with the nature and clues of the Hamiltonian in the early universe that is Cotton–York dominated.
 The details of the origin of the universe that ITG pieces together will soon be discussed\footnote{Notes on gravitational Chern–Simons functional, Cotton–York tensor, and Euclidean Pontryagin invariant which may be helpful to understanding the arguments are delegated to Appendix \ref{AB}.}; it advocates the Chern–Simons Hartle–Hawking initial state as the quantum origin of our universe.

\subsection{Semiclassical  Robertson–Walker Beginning}

A noteworthy feature of ITG is that Euclidean and Lorentzian four-manifolds can interface at a common spatial volume, i.e., at a particular instant of intrinsic time, provided junction conditions for signature change are met.
A requirement of vanishing momentum at the interface suffices \cite{Hayward}.  As the commutator $[{\bar\pi}^i_j , {\tilde C}^j_i ]$ vanishes identically, the Hamiltonian of ITG can be reexpressed as
\begin{eqnarray}
\label{HAMIL1}
H &=&\frac{1}{\beta}\int \sqrt{\bar{\pi}^i_j\bar{\pi}^j_i - \frac{q}{(2\kappa)^2}(R-2\Lambda)+g^2\hbar^2\tilde{C}^i_j\tilde{C}^j_i} \,d^3x  \nonumber \\
&=& \frac{1}{\beta}\int \sqrt{ (Q^\dagger_{CS})^i_j(Q_{CS})^j_i - \frac{q}{(2\kappa)^2}(R-2\Lambda)}\, d^3x;  \\
 (Q_{CS})^i_j &:=& e^{-gW_{CS}}{\bar\pi}^i_j  e^{gW_{CS}} = {\bar\pi}^i_j  - i g\hbar{\tilde C}^i_j. \nonumber
\end{eqnarray}
The null expectation value of the complex combination $(Q_{CS}) ^i_j $ implies the classical vanishing of both the traceless momentric $\bar{\pi}^i_j$ and the Cotton–York tensor density $\tilde{C}^i_j$.  The latter is precisely the statement of spatial conformal flatness.
Since the Hamiltonian density in ITG  is proportional to ${\tilde\pi}$, the remaining criterion of the junction condition is that spatial Ricci scalar $R$ cancels the effective cosmological constant term on the conformally flat spatial interface with zero momentric (see also Appendix \ref{AA} on the Lichnerowicz–York equation for a complementary viewpoint). For Einstein's theory and also for ITG, the vanishing of total momentum and Cotton–York tensor is the initial data of a Lorentzian de Sitter solution with $k=+1$ conformally flat $S^3$ slicings\footnote{It is known that every compact simply connected conformally flat manifold is conformally diffeomorphic to the $n$-sphere \cite{Kuiper}. Grigori Perelman proved the Poincar\'{e} Conjecture, which says that every simply connected, closed three-manifold is homeomorphic to the three-sphere.  Topological and differentiable isomorphism is the same in dimension three and below, so the generalized smooth Poincar\'{e} Conjecture that a manifold which is a homotopy three-sphere is, in fact, diffeomorphic to the standard three-sphere holds.}; and the Lorentzian-Euclidean interface happens precisely at the throat of the Lorentzian de Sitter solution. Continuation to purely Euclidean signature manifold (which is a hemisphere of $S^4$) occurs at smaller volumes or earlier intrinsic times (see Figure \ref{fig1}); at the interface, which is the equator of the Euclidean hemisphere as well as the de Sitter Lorentzian throat, $R =6k/a^2 = 2\Lambda$.

This Euclidean–Lorentzian continuation is an exact and explicit instanton tunneling solution of the emergence of a Lorentzian signature universe from a Euclidean origin of a single point, which Hawking had presented \cite{Nature of spacetime}. ITG  bolsters this imagery, and it finds correspondence and happy union between the Hartle–Hawking Proposal and Penrose's Weyl Curvature Hypothesis.
This picture is the semiclassical result of an exact quantum wave function, $\Psi_{CS} ={N'}\exp{(-gW_{CS})}$, which is the exponent of the Chern–Simons functional of the affine connection $W_{CS}$ (see Appendix \ref{AB}).
This Chern–Simons Hartle–Hawking state is the solution of $(Q_{CS})^i_j|\Psi_{CS}\rangle =0$\footnote{Explicitly, $(Q_{CS})^i_j \Psi_{CS}= (e^{-gW_{CS}}{\bar\pi}^i_j  e^{gW_{CS}} )\Psi_{CS} ={e^{-gW_{CS}}{\bar\pi}^i_j {N'}}=0$, provided ${N'}$ is a topological invariant which satisfies $\frac{\delta {N'}}{\delta {\bar q}_{ij}} =0$. In the metric representation, the self-adjoint momentric operator can be expressed as \cite{ITGBook}
$\bar{\pi}_{i}^{j} =\frac{\hbar}{i}(\frac{1}{2}\bigl(\delta^i_m\bar{q}_{jn}+\delta^i_n\bar{q}_{jm}\bigr)-\frac{1}{3}\delta^i_j\bar{q}_{mn})\frac{\delta}{\delta \bar q_{mn}}$.}
; its critical point $\frac{\delta W_{CS}}{\delta q_{ij}}= {\tilde C}^{ij}=0$ corresponds to conformal flatness; and the semiclassical momentum ${\tilde\pi}^{ij} \propto
\frac{\delta W_{CS}}{\delta q_{ij}} = {\tilde C}^{ij}$ vanishes at this critical value.  The correspondence goes beyond critical point analysis.
For this state, $\langle\Psi_{CS}|(Q_{CS})^i_j|\Psi_{CS}\rangle =0$ implies
$\langle\Psi_{CS}|{\bar\pi}^i_j|\Psi_{CS}\rangle =0$ and $\langle\Psi_{CS}|{\tilde C}^i_j|\Psi_{CS}\rangle =0$ actually hold at the level of expectation values\footnote{This is analogous to the situation in the simple harmonic oscillator in which the action of the complex annihilation operator, ${a}$, on the vacuum state yields $\langle 0|{a}|0\rangle =0$, consequently  $\langle 0|x|0\rangle = \langle 0|p|0\rangle =0$.}. To wit, with spatial $S^3$ conformal flatness and vanishing momentric as classical values, it can be demonstrated, as in Refs. \cite{ITGBook}, that the equations of motion of the Hamiltonian \eqref{HAMIL1} of ITG evolves this set of initial data into the Lorentzian de Sitter manifold with $k=+1$ conformally flat $S^3$ spatial slicings. The corresponding metric can be expressed as
\begin{equation}
ds^2 =-dt^2 +\frac{3}{\Lambda}cosh^2[\sqrt{\frac{\Lambda}{3}}(t-t_{throat})]dl^2_{S^3}.
\end{equation}
This explicit Lorentzian de Sitter solution has a minimum spatial volume at the throat. At even smaller volumes, analytic continuation to purely Euclidean signature (with Euclidean time interval $d\tau =-idt$)  yields the metric,
 \begin{equation}
 \label{Euclideanperiod}
ds^2 =d\tau^2 +\frac{3}{\Lambda}cos^2[\sqrt{\frac{\Lambda}{3}}(\tau-\tau_{throat})]dl^2_{S^3}.
\end{equation}
This continuation constitutes a hemisphere of Euclidean de Sitter $S^4$ manifold (see Figure \ref{fig1}), which is, in fact, a Riemannian Einstein manifold, as well as a solution (as shown in Chapters 6 and 10 of Ref. \cite{ITGBook}) of the corresponding equations of motion of the Hamiltonian of ITG\footnote{It is a theorem \cite{GRWaves} that a solution of Einstein's theory with conformally flat (i.e., ${\tilde C}^i_j=0$)  spatial slicings is also a solution of the equations of motion of ITG since the two theories differ by only a square term ${\tilde C}^i_j {\tilde C}^j_i$ in ${\mathcal V}$ of \eqref{Hbar}.}. This Euclidean–Lorentzian continuation is an exact and explicit example of instanton tunneling from a Euclidean origin at a point of zero volume to a Lorentzian de Sitter universe of finite spatial size.
Gravitational instantons are non-singular solutions of classical Einstein equations on some section of complexified spacetime \cite{HawkingGibbonsInstanton1,HawkingGibbonsInstanton2}; an even closer analogy to Yang–Mills instantons happens when these gravitational solutions also exhibit self-dual Riemannian or Weyl curvature two-forms. This is true of the  $S^4$  Euclidean de Sitter manifold, which is conformally flat in four dimensions (ergo, it is self-dual in Weyl curvature, albeit trivially so).
Thus, for the explicit de Sitter solution, Penrose's classical criterion of vanishing Weyl curvature actually holds for the initial data of vanishing momentum and Cotton–York tensor at the beginning of the universe (the de Sitter solution is a special case of Robertson–Walker spacetimes, all of which come with vanishing Weyl tensor).

\subsection{A Hot Beginning}

The current CMB radiation with temperature variations of one part in $10^5$ is a very good approximation of a black body spectrum, and it is even closer to the Planck spectrum at earlier times before structure formation.
Emergence from Euclidean tunneling offers a tantalizing explanation of a hot beginning: The ``Big Bang'' (the term is used here to refer to the beginning of the Lorentzian universe from a tiny size rather than from a point of infinite curvature, density, and temperature)
happens at the Euclidean–Lorentzian interface, which has a Euclidean time variable that is periodic.
The explicit Lorentzian de Sitter solution is joined at the throat by the Euclidean de Sitter manifold \eqref{Euclideanperiod}, which exhibits a periodic time variable. The scale factor at the throat is $a = \sqrt{\frac{3}{\Lambda}}$, and the periodicity of the Euclidean time variable $\tau$ governed by the effective value of the cosmological constant is ${\cal P}=\frac{2\pi}{c}\sqrt{\frac{3}{\Lambda}}$. The upshot is that correlation functions of fields should obey the Kubo-Martin-Schwinger condition~\cite{KMS1,KMS2,KMS3}, which is a manifestation in statistical mechanics of quantum mechanical systems and quantum field theory of a system in thermal equilibrium, with the partition function of a thermal system at temperature ${\cal T}_{throat} =\frac{\hbar}{k_B{\cal P}}$.
 A Planck scale interface with $a \sim L_P$ implies  ${\cal T} =\frac{\hbar c}{2\pi k_B L_P} =\frac{1}{2\pi}{\cal T}_P\sim 10^{31}$K. A very hot beginning is associated with the narrowness of the throat and the high value of effective cosmological constant at that instant.

\section{Quantum Origin as Chern–Simons Hartle–Hawking State}

As the exterior derivative of the Chern–Simons three-form is the Pontryagin integrand, $W_{CS}[\Gamma|_{\Sigma=\partial M}] =-\frac{1}{2}\int_M tr(R\wedge R)$, a Chern–Simons state $e^{-gW_{CS}[\Gamma_{\Sigma}]} =e^{\frac{g}{2}\int_M tr(R\wedge R)}$  is in fact expressible as the boundary contribution of the exponential of a Euclidean action which is the Pontryagin integral.
However, the Chern–Simons functional is not invariant under large gauge transformations, so two generic compact Euclidean manifolds, $M$ and $M'$, with the same boundary $\Sigma$ may yield the same Chern–Simons contribution up to the winding number of a large gauge transformation. As the Chern–Simons state is a real exponential factor, rather than a mere phase, the winding number discrepancy cannot be modded
out by quantizing the coupling constant $g$ (as in theories with three-dimensional Chern–Simons {\it action}). Furthermore, the real exponential factor will be unbounded as it can be arbitrarily stepped up or down with the number of windings. However, there is a device to define the Chern–Simons functional modulo large gauge transformations. This can be done by subtracting a topological invariant that is the Chern–Simons functional of a flat connection in Yang–Mills gauge theory \cite{ITGBook}, or, of a conformally flat connection, $\Gamma_o|_\Sigma$, for the metric gravitational case at hand. Since Chern–Simons functional  $W_{CS}[\Gamma_o|_\Sigma]$ changes by the same winding number, the difference
 \begin{equation}
 \label{CSmodulo}
W_{CS}[\Gamma'|_\Sigma]-W_{CS}[\Gamma'_o|_\Sigma]=W_{CS}[\Gamma|_\Sigma]-W_{CS}[\Gamma_o|_\Sigma],
\end{equation}
 is invariant under $U$  for $\Gamma'= U\Gamma U^{-1} + UdU^{-1}, \Gamma'_o= U\Gamma_o U^{-1} + UdU^{-1}$ on $\Sigma=\partial M=\partial M'$, regardless of the winding number of $U$.
 This difference can be expressed as
 \begin{equation}
W_{CS}[\Gamma|_\Sigma]-W_{CS}[\Gamma_o|_\Sigma]=-\frac{1}{2}\int_{M} tr({R\wedge R}) -W_{CS}[\Gamma_o|_{\Sigma=\partial M}].
\end{equation}
So, a Hartle–Hawking proposal with Pontryagin Euclidean action, modified by subtraction of a conformally flat Chern–Simons functional at the boundary, leads to the state
 \begin{equation}
\Psi_{\rm{Hartle-Hawking}}[\Gamma|_\Sigma] =N\int D\Gamma_{M}\,  e^{-g(-\frac{1}{2}\int_{M:\partial{M}=\Sigma} tr(R\wedge R) -W_{CS}[\Gamma_o|_{\Sigma=\partial M}])},
\end{equation}
wherein we sum over {\it all} compact four-manifolds $M$ with the criterion the affine connection reduces to  $\Gamma$ on $\Sigma=\partial M$.
The corresponding Hartle–Hawking wave function is, therefore,
\begin{equation}\label{psinoboundary}
\Psi_{\rm{Hartle-Hawking}}[\Gamma|_\Sigma]={\cal N}e^{-g(W_{CS}[\Gamma(q_{ij})|_\Sigma] -W_{CS}[\Gamma_o|_\Sigma])}.
\end{equation}
This, too, has the interpretation as the amplitude for a three-geometry to arise from a single point.
Since $\Sigma$ is the boundary of Euclidean signature manifolds, it follows that $\Sigma$ will naturally and advantageously be Riemannian rather than pseudo-Riemannian; furthermore, $\Sigma$, being the boundary of a compact (four) manifold, must be closed.

\subsection*{Relative Chern–Simons Functional}
The wave function of the initial state can be stated in the more rigorous mathematical language \cite{Moore} of relative Chern–Simons forms ${CS}[\Gamma_2, \Gamma_1]$.
Using a connection $\Gamma(s)= (1-s)\Gamma_1 + s\Gamma_2 =: \Gamma_1 +s \Delta\Gamma$  (for $0 \leq s \leq 1$)  which interpolates between $\Gamma_1$ and  $\Gamma_2$, and the Bianchi identity $D^{\Gamma(s)}R(s)=0$,
a particular application of Chern-Weil theory yields the difference,
\begin{eqnarray} \label{RCS}
 tr(R_2\wedge R_2)&-&tr(R_1\wedge R_1)=\int^1_{s=0} ds\frac{d}{ds} tr(R(s)\wedge R(s))\nonumber\\
 &=& \int^1_{s=0} ds\, tr(2(D^{\Gamma(s)}\Delta\Gamma)\wedge R(s))\nonumber\\
 &=&d( 2\int^1_{s=0} ds\, tr(\Delta\Gamma\wedge R(s)))\nonumber\\
 &=&d( 2\int^1_{s=0} ds\, tr(\Delta\Gamma\wedge(R_1 +sD^{\Gamma_1} \Delta\Gamma + s^2 \Delta\Gamma\wedge\Delta\Gamma)))\nonumber\\
 &=& d{CS}(\Gamma_2, \Gamma_1),
\end{eqnarray}
\begin{equation}
 {CS}(\Gamma_2, \Gamma_1):=tr(2\Delta\Gamma\wedge R_1 +\Delta\Gamma\wedge D^{\Gamma_1}\Delta\Gamma +\frac{2}{3} \Delta\Gamma\wedge\Delta\Gamma\wedge\Delta\Gamma).
 \end{equation}
 As ${CS}(\Gamma_2, \Gamma_1)$ involves only tensorial quantities, this difference is {\it globally} exact and well defined. $\Delta\Gamma=\Gamma -\Gamma_o$ transforms covariantly as the difference between two connections (rather than inhomogeneously as a connection);
 and the relative Chern–Simons form is invariant, $CS(\Gamma^U_2,\Gamma^U_1)=CS(\Gamma_2,\Gamma_1)$, under $\Gamma^U_{1,2}= U\Gamma_{1,2}U^{-1} + UdU^{-1} $ for all $U$, including ``large gauge transformations'' with nontrivial winding numbers.
 This is true, in particular, of the Chern–Simons functional $W_{CS}[\Gamma, \Gamma_o] :=-\frac{1}{2}\int_\Sigma CS[\Gamma,\Gamma_o]$ of $\Gamma$ relative to a conformally flat connection  $\Gamma_o$.

 The  Chern–Simons Hartle–Hawking wave function discussed earlier can thus be phrased in the precise context of the relative Chern–Simons functional that is
\begin{eqnarray}\label{HHCSstate}
\Psi_{CS}=\Psi_{\rm{Hartle-Hawking}} &=& {\cal N}\exp(-gW_{CS}[\Gamma, \Gamma_o]); \\
W_{CS}[\Gamma, \Gamma_o] &:=&-\frac{1}{2}\int_\Sigma CS[\Gamma,\Gamma_o]. \nonumber
\end{eqnarray}
 The subtraction of the Chern–Simons functional of a conformally flat connection does not alter the fact that this relative Chern–Simons state satisfies
$(Q_{CS})^i_j \Psi_{CS}=0$. This is because $[{\bar\pi}^i_j, W_{CS}[\Gamma_o]]=({\tilde C}|_o)^i_j  =0$, with $({\tilde C}|_o)^i_j $ denoting the Cotton–York tensor density of a conformally flat connection (as discussed, the Cotton–York tensor vanishes iff there is conformal flatness).

\section{Quantum Metric Fluctuations and Two-Point Correlation Functions of the Chern–Simons Hartle–Hawking State}

The Chern–Simons Hartle–Hawking state, as explained, gives rise to zero expectation values for the momentric and Cotton–York tensor. As well as semiclassical description,
an exact quantum state contains fluctuations beyond the mean values, and these quantum correlations are salient features and defining characteristics of the state.

With the relative Chern–Simons functional, the quadratic term of an expansion about the $S^3$ background results in\vspace{-12pt}
 \begin{equation}
\int d^{3}x\int d^{3}y\,\Delta{\bar q}_{ij}(x)\left.\frac{\delta^2 W_{CS}}{\delta \bar{q}_{ij}(x) \delta \bar{q}_{kl}(y)}\right|_{S^3}\Delta{\bar q}_{kl}(y)
=: \int d^{3}x\int d^{3}y\, \Delta{\bar q}_{ij}(x) H^{ijkl}(x,y)\Delta{\bar q}_{kl}(y).
\end{equation}
The critical point of the Chern–Simons functional is precisely at the vanishing of the Cotton–York tensor, i.e., at conformal flatness. For the simply connected spatially closed case, this is uniquely the standard three-sphere $S^3$, which is chosen as the background above.
About $S^3$, with unimodular metric changes restricted to physical TT infinitesimal changes $\delta{\bar q}^{TT}_{ij}$ (the TT excitations are orthogonal to Lie derivative changes in ${\bar q}_{ij}$)  yields the Hessian as
 \begin{equation}
\left.H^{ijkl}(x,y)\right|_{S^3}=
\left.\frac{1}{8}((\bar{q}^{ik}\tilde\epsilon^{jlm}+\bar{q}^{il}\tilde\epsilon^{jkm}+\bar{q}^{jk}\tilde\epsilon^{ilm}+\bar{q}^{jl}\tilde\epsilon^{ikm})\bar{\nabla}_{m}\bar{\nabla}^{2})\right|_{S^3}\delta^{3}(x,y).
\label{Hessian}
\end{equation}
The inverse operator is
\begin{equation}
\left.H^{-1}_{ijkl}(x,y)\right|_{S^3}=
\left.\frac{1}{2}((\bar{q}_{ik}\epss_{jrl}+\bar{q}_{il}\epss_{jrk}+\bar{q}_{jk}\epss_{irl}
+\bar{q}_{jl}\epss_{irk})\frac{1}{\bar{\nabla}^2}\bar{\nabla}^{r}\frac{1}{\bar{\nabla}^2})\right|_{S^3}\delta^{3}(x,y).
\label{HinverseH}
\end{equation}


Expectation values of operators with respect to the Chern–Simons Hartle–Hawking state can be evaluated explicitly.
The two-point correlation function for TT modes is defined by\footnote{
$|\Psi_{CS}\rangle$ has a norm which is essentially just the partition function of the relative Chern–Simons action,
$\langle\Psi_{CS}|\Psi_{CS}\rangle \propto Z = \int \mathfrak{D}\overline{q}\, e^{-2g (W_{CS}[\Gamma]-W_{CS}[\Gamma_o])}$.}
\begin{equation}\langle\Psi_{CS}|\delta{\bar q}^{TT}_{ij}(x)\delta{\bar q}^{TT}_{kl}(y)|\Psi_{CS}\rangle
=  \frac{1}{Z}\int \mathfrak{D}\bar{q}\,\, \delta{\bar q}^{TT}_{ij}(x)\delta{\bar q}^{TT}_{kl}(y) e^{-2g (W_{CS}[\Gamma]-W_{CS}[\Gamma_o])}.
\end{equation}
Expanding the exponent in terms of the coupling constant\footnote{In usual practice, $g$ is absorbed in the quadratic ``propagator'' term by redefinition, and $\frac{1}{\sqrt g}$ is considered to be ``the coupling constant'' in the interaction terms.
So, in usual QFT language, a large value for $g$ implies weak coupling, and vice versa.}, and
noting that the inverse of the above Hessian is essentially the two-point function, to lowest-order approximation,
\begin{equation}
\label{qttcorrelation}
\langle\Psi_{CS}|\delta{\bar q}^{TT}_{ij}(x)\delta{\bar q}^{TT}_{kl}(y)|\Psi_{CS}\rangle=\frac{1}{2g}\left.H^{-1}_{ijkl}(x,y)\right|_{S^3}.
\end{equation}
With flat coordinates for $\bar{\nabla}_i$, the inverse Hessian of  Equation \eqref{HinverseH} just works out in momentum space, or ``$k$-space'',  to be ``$\frac{1}{k^3}$  in character''.
This is the well-known signature of a power spectrum of scale-invariant perturbations. Explicitly for \eqref{HinverseH}, with conformally flat background (hence flat ${\bar q}_{ij} =\delta_{ij}$),
\be
\label{Fourierk3}
\langle\Psi_{CS}|\delta{\bar q}^{TT}_{ij}({\bf x})\delta{\bar q}^{TT}_{kl}({\bf y})|\Psi_{CS}\rangle\propto \int d^3k (\bar{q}_{ik}\epss_{jrl}+\bar{q}_{il}\epss_{jrk}+\bar{q}_{jk}\epss_{irl}
+\overline{q}_{jl}\epss_{irk}) \frac{k^r}{k^4}e^{i{\bf k}\cdot({\bf x}-{\bf y})};
\en
remains unchanged under ${\bf x}\rightarrow  s{\bf x}, {\bf y}\rightarrow s{\bf y}$ scaling (by redefining the dummy variable ${\bf k} \rightarrow \frac{1}{s}{\bf k}$ in the integration).

The $S^3$ spatial manifold is closed; fortuitously, the explicit TT modes and their eigenvalues are known exactly \cite{Lindblom} even when flat space Fourier decomposition is not applicable.
Every physical $\delta{\bar q}^{TT}_{ij}$ can be expanded in terms of these orthonormal modes as
\begin{equation}
\delta{\bar q}^{TT}_{ij} =\sum_{\{K\}} c^{+}_{\{K\}}{\bar Y}^{\{K\}}_{(+)ij} + c^{-}_{\{K\}}{\bar Y}^{\{K\}}_{(-)ij};
\end{equation}
wherein ${\bar Y}^{\{K\}}_{(\pm)ij}$ are chosen to be orthonormal modes that diagonalize the Hessian.
These TT modes are described explicitly in Ref. \cite{ITGBook}, and they are adapted from the TT modes of $S^3$ derived and listed in Ref. \cite{Lindblom}; $\{K\}$ denotes the full set of quantum numbers of the modes, $\{Klm\}$, in these references.
With these modes the eigenvalues (there are two eigenvalues which are opposite in value as the Hessian is traceless) can be computed explicitly.  To wit, the quadratic term yields
 \begin{equation}
 \int d^{3}x\int d^{3}y\,\delta{\bar q}^{TT}_{ij}(x) H^{ijkl}(x,y)\delta{\bar q}^{TT}_{kl}(y)
=\sum_{\{K\}}\frac{{[K(K+2)-2]}^{\frac{3}{2}}}{2a^3}((c^+_{\{K\}})^* c^+_{\{K\}} - (c^-_{\{K\}})^* c^-_{\{K\}}).
\end{equation}
The correlation function $\langle\Psi_{CS}|\delta{\bar q}^{TT}_{ij}(x)\delta{\bar q}^{TT}_{kl}(y)|\Psi_{CS}\rangle$ can be expressed through the fundamental two-point functions of these eigenmodes.
At this level of approximation, the $\{K\}$-space behavior of the correlation functions is thus \cite{ITGBook}
\begin{eqnarray}
\label{aacorrelation}
\langle\Psi_{CS}|(c^{\pm}_{\{K\}})^*c^{\pm}_{\{K'\}}|\Psi_{CS}\rangle &=& \frac{2a^3}{g{[K(K+2)-2]}^{\frac{3}{2}}}\delta_{\{K\}\{K'\}}, \\
\langle\Psi_{CS}|(c^{\pm}_{\{K\}})^*c^{\mp}_{\{K'\}}|\Psi_{CS}\rangle &=&0;
\end{eqnarray}
wherein $a$ is the $k=+1$ Roberston-Walker scale factor of the three-sphere.

The dependence of the correlation functions can be stated more succinctly in terms of the eigenvalues of the Laplacian operator.
These modes satisfy \cite{Lindblom}
\begin{eqnarray}
{\bar\nabla}^2 {\bar Y}^{Klm}_{(4,5)ij} &=& -\lambda_K{\bar Y}^{Klm}_{(4,5)ij}, \nonumber \\
\lambda_K &=& \frac{K(K+2)-2}{a^2}\quad \rm{with}\,\, K\geq 2,
\end{eqnarray}
Therefore $\langle\Psi_{CS}|(c^{\pm}_{\{K\}})^*c^{\pm}_{\{K'\}}|\Psi_{CS}\rangle$ in Equation \eqref{aacorrelation} varies precisely as $(\lambda_K )^{-3/2}$.
This is in complete agreement with the scale-invariant $k^{-3}$ dependence in the earlier Fourier analysis in Equation \eqref{Fourierk3} with flat coordinates. Therein the eigenvalues of ${\bar\nabla}^2$ for Fourier modes are $-k^2$.
Expressed in terms of coordinate independent eigenvalues of the Laplacian, the $(\lambda_K )^{-3/2}$ dependence is thus the generalized characterization of ``scale-invariant correlations''.

 Variations of the Cotton–York tensor are one-to-one with variations of $\delta{\bar q}^{TT}_{ij}$ when the Hessian is invertible.
 The Cotton–York tensor, which is also TT, may be regarded as the non-perturbative manifestation of the two physical d.o.f. in geometrodynamics.
 So the physical excitations $\delta{\bar q}^{TT}_{ij}$  translates directly into Cotton–York tensor curvature perturbations $\delta{\tilde C}^i_j (x) = \int \left.H^i\,_j\,^{kl}(x,y)\right|_{S^3}\delta{\bar q}^{TT}_{kl} (y) d^{3}y$
 about conformally flat $S^3$. From this and Equation \eqref{qttcorrelation}, the two-point correlation function of the Chern–Simons state for Cotton–York fluctuations about the conformally flat $S^3$ saddle point is
 \begin{equation}
 \langle\Psi_{CS}|\delta{\tilde C}^i_j (x)\delta{\tilde C}^k_l (y)|\Psi_{CS}\rangle = \frac{1}{2g}\left.H^i\,_j\,^k\,_l(x,y)\right|_{S^3}.
 \end{equation}

 A signature of the Chern–Simons Hartle–Hawking origin of the universe is that it manifests, at the lowest-order approximation, scale-invariant two-point correlations for transverse traceless quantum metric fluctuations. In the correlation functions, no less than metric and curvature fluctuations in quantum gravity are being manifested. The defining characteristics of these correlations emerge from the culmination and confluence of the Hartle–Hawking proposal, the Chern–Simons state, and the framework of Intrinsic Time Geometrodynamics.



\vspace{6pt}

\acknowledgments{I am thankful for the many stimulating discussions with Hoi-Lai Yu on physics, life, and the universe.}

\clearpage

\appendix
\section[\appendixname~\thesection]{Lichnerowicz–York Equation with Addition of Cotton–York Term and Penrose's Weyl Curvature Hypothesis}\label{AA}


The initial data problem in Einstein's theory has been solved by York in a formulation \cite{extrinsic,York1,York2,Niall1,Niall2}, which casts the Hamiltonian constraint in the form of the Lichnerowicz–York equation. The equation determines the conformal factor $\phi$.
It is interesting to investigate within this classical context the implications that arise with the addition of a Cotton–York term to Einstein's theory \cite{SOOYU1,ITGBook}. To wit, the corresponding Hamiltonian constraint with the York term is
\begin{equation}\label{H6}
-\alpha^2q(R-2\Lambda) + \bar{q}_{ik}\bar{q}_{jl}{\bar\pi}^{ij}{\bar\pi}^{kl} +g^2 {\tilde C}^i_j{\tilde C}^j_i- \beta^2{\tilde\pi}^2 = 0.
\end{equation}
With the Cotton–York addition, the generalized Lichnerowicz–York equation, which corresponds to the constraint, becomes
\begin{equation}\label{LY1}
8\alpha^2\bar{\nabla}^2\phi - \alpha^2(\bar{R}\phi -2\Lambda\phi^5) + {q}^{-1}({\tilde\pi}^{iTT}_j{\tilde\pi}^{jTT}_i + g^2{\tilde C}^i_j{\tilde C}^j_i)\phi^{-7}
 = \beta^2\hat{p}^2\phi^5,
\end{equation}
wherein ${\bar R}$ is the scalar curvature of the unimodular spatial metric ${\bar q}_{ij}$, $\phi^4 :=q^{1/3}$, and $\alpha^2 := \frac{1}{(2\kappa)^2}$.
In York's formulation, $\hat p := \tilde\pi/{\sqrt q}$ is chosen to be spatially independent, and together with transverse traceless ${\tilde\pi}^{iTT}_j$,  the momentum constraint, $-2\nabla_j{\tilde\pi}^j_i =0$, is solved with ${\tilde\pi}^j_i :={\tilde\pi}^{jTT}_i +\frac{1}{3}\sqrt{q}\delta^j_i{\hat p}$.
It is worth emphasizing the addition of Cotton–York and other higher three-curvature terms to the potential explicitly breaks the four-covariance of Einstein's theory. Despite this, the initial data formulation (with transverse momentum and solving the Hamiltonian constraint with the Lichnerowicz–York equation) remains valid, which proves that we can consistently depart from Einstein's theory, keeping three-covariance and maintaining the same two d.o.f. as in GR.

We can compare how each term scales with $q$ in the equation. The Cotton–York tensor density $\tilde C^{i}_{j}$ is conformally invariant, ergo $q$-independent.
With the modified Lichnerowicz–York Equation \eqref{LY1},  at late intrinsic times, i.e., large spatial volumes or large values of $\phi$, compared to the scalar curvature and cosmological constant terms, the Cotton–York contribution vanishes rapidly with $\phi^{-7}$ multiplying it;
therefore, Einstein's theory dominates, and the Cotton–York addition and its attendant physical effects will be suppressed.  However at very small volumes at the beginning with $\phi \rightarrow 0$,  the divergence in the equation can be tamed only if ${\tilde\pi}^{iTT}_j{\tilde\pi}^{jTT}_i$  and the Cotton–York term,  ${\tilde C}^i_j{\tilde C}^j_i$, {\it both} vanish. Although the vanishing of the traceless part of the momentum is a statement on the extrinsic curvature, it says nothing of the intrinsic spatial geometry. This poses less restriction on the initial singularity of the universe (for instance, as Penrose has pointed out, black hole singularities have divergent Weyl tensors, whereas Robertson–Walker spacetimes have four-dimensional Ricci scalar singularity and vanishing Weyl curvature). In contradistinction, the new ingredient of the vanishing Cotton–York tensor is the precise requirement of spatial conformal flatness on the intrinsic geometry. The only simply connected closed conformally flat spatial manifold is the standard $S^3$. Together with vanishing ${\tilde\pi}^{iTT}_j$, the non-triviality only resides in $\phi$ and ${\hat p}$.
To wit, with ${\tilde C}^i_j ={\tilde\pi}^{iTT}_j=0$, the equation reduces to
\be
-\phi{\bar R} + 8\bar{\nabla}^2\phi + 2\Lambda\phi^5  = \frac{\beta^2}{\alpha^2 }\hat{p}^2\phi^5.
\en
A standard sphere of radius $a$ yields $R =6/a^2= \phi^{-5}(\phi{\bar R} -8{\bar\nabla}^2\phi )$. This implies $\hat{p}^2 = \frac{6\alpha^2}{\beta^2} (\frac{\Lambda}{3} - \frac{1}{a^2})$,
which is a constant mean curvature (CMC) slice, and the remaining equation,
\be\label{phiequation}
 -{\bar\nabla}^2\phi +\frac{\phi{\bar R}}{8}=\frac{3}{4a^2}\phi^5,
\en
is still non-linear in flat coordinates with ${\bar R} =0$ (since $R$ is conformally flat), but it has\linebreak $\phi = \sqrt{\frac{2a}{1+r^2}}$ as exact solution. This is the expression of Robertson–Walker metric in conformally flat spatial coordinates,
\be \label{conformalRW}
ds^2 = -dt^2  + \frac{4a^2}{(1+r^2)^2}(dr^2 +r^2(d\theta^2 + \sin^2\theta d\phi^2)).
\en
The new Cotton–York ingredient gives credence to Penrose's Hypothesis of initial vanishing Weyl curvature tensor and smooth Robertson–Walker beginning, whereas Einstein's theory, {\it sans} Cotton–York addition, does not single out our ``extraordinarily special'' universe.

It is also noteworthy that an initial data set of conformal flatness and vanishing ${\pi}^{iTT}_j$ together with CMC  is compatible with the initial data for Lorentzian de Sitter solution with $S^3$ slicings.
$\tilde\pi$ vanishes at the de Sitter throat at which ${\hat p}=0$  i.e., at $a =\sqrt{\frac{3}{\Lambda}}$). The complete vanishing of all momentum or extrinsic curvature at the throat is, moreover, the correct junction condition \cite{Hayward} for Euclidean–Lorentzian continuation.
The York initial data formulation is classical in context, but remarkably, an exact Chern–Simons Hartle–Hawking quantum wave function (as described earlier in the article proper) can be found, with conformal flatness and vanishing momentric as expectation values.

\section[\appendixname~\thesection]{Gravitational Chern–Simons Functional, Cotton–York Tensor, and Pontryagin Invariant }\label{AB}
The Chern-Simon functional \cite{Chern-Simons} of the spatial metric affine connection is
\begin{equation}
\begin{aligned}
W_{CS}&=\frac{1}{2}\int {\tilde\epsilon}^{ikj} ( \Gamma^{m}_{in}\partial_{j}\Gamma^{n}_{km}+\frac{2}{3}\Gamma^{m}_{in}\Gamma^{n}_{js}\Gamma^{s}_{km} )d^{3}x.
\label{W}
\end{aligned}
\end{equation}
Its variation leads, upon integration by parts over closed spatial manifold, to $\delta W_{CS}=\frac{1}{2}\int {\tilde\epsilon}^{ikj}R^{n}\,_{mjk}\delta\Gamma^{m}_{in} \, d^{3}x$; while the variation of the connection results in $\delta\Gamma^{m}_{in}=\frac{1}{2}q^{mr}(\nabla_{i}\delta q_{rn}+\nabla_{n}\delta q_{ri}-\nabla_{r}\delta q_{in})$.  In three dimensions, the Riemann curvature tensor
can be expressed completely in terms of the Ricci tensor through
\begin{equation}
\begin{aligned}
R^{n}\,_{mjk}=q_{mk}R^{n}_{j}+\delta^{n}_{j}R_{mk}-q_{mj}R^{n}_{k}-\delta^{n}_{k}R_{mj}+\frac{R}{2}(\delta^{n}_{k}q_{mj}-\delta^{n}_{j}q_{mk}).
\label{RRc}
\end{aligned}
\end{equation}
These results conspire to yield
\begin{equation}
\delta W_{CS}=\int {\tilde\epsilon}^{imn}\nabla_{m}(R^{j}_{n}-\frac{1}{4}\delta^{j}_{n}R)\delta q_{ij}d^{3}x;
\label{vW3}
\end{equation}
thus, the functional derivative of the Chern–Simons functional $W_{CS}$ with respect to the spatial metric yields the Cotton–York tensor density \cite{York-Cotton1,York-Cotton2} i.e.,
\begin{equation}
\frac{\delta W_{CS}}{\delta q_{ij}}= {\tilde C}^{ij} :=  {\tilde\epsilon}^{ikl}\nabla_k (R^j\,_{l}-\frac{1}{4}\delta^{j}\,_lR).
\end{equation}
This entity differs saliently in being a third-order derivative of the spatial metric, whereas Yang–Mills and Riemann curvatures are, respectively, first-order and second-order derivatives of the connection and the metric; ergo ${\tilde C}^{ij}$ has physical dimension (length)$^{-3}$ instead of (length)$^{-2}$. The antisymmetric part of ${\tilde C}^{ij}$ vanishes via the Bianchi identity. Therefore, it can also be written in explicitly symmetric traceless form
  \begin{eqnarray}
{\tilde C}^{ij} &=&{\tilde\epsilon}^{imn}\nabla_m (R^j\,_{n}-\frac{1}{4}\delta^{j}\,_nR) \nonumber\\
  &=&  \frac{1}{2}({\tilde\epsilon}^{imn}\nabla_m R^j\,_{n} + {\tilde\epsilon}^{jmn}\nabla_mR^i\,_{n}).
  \end{eqnarray}

 $W_{CS}$ possesses the additional property of being dependent only on the unimodular part, ${\bar q}_{ij}$, of the spatial metric ($q_{ij}= q^{\frac{1}{3}}{\bar q}_{ij})$ and independent of $q$  i.e., $W_{CS}[\Gamma^i_{jk}(q_{kl})] = W_{CS}[{\bar\Gamma}^i_{jk}({\bar q}_{kl})]$.  It is invariant under spatial diffeomorphisms of the metric, as $\delta W_{CS}=\int {\tilde C}^{ij}{\delta q}_{ij}\,d^3x$ vanishes (due to the transversality and symmetry of the Cotton–York tensor, and upon integration by parts) when the metric changes under spatial diffeomorphisms by ${\delta q}_{ij} = {\cal L}_{\vec N}q_{ij} = \nabla_i N_j + \nabla _j N_i$.

In four-dimensional manifolds with Euclidean signature, the Pontryagin invariant is $P_4 = \frac {1}{8\pi^2}\int_M tr(R\wedge R)$  with $R= d\Gamma +\Gamma\wedge\Gamma$ being the curvature two-form.
Since
\begin{equation}
  \int_{M} tr(R\wedge R) = \int_{M} d (tr(\Gamma d\Gamma +\frac{2}{3}\Gamma \wedge\Gamma \wedge\Gamma)),
\end{equation}
if a three-manifold of interest, $\Sigma$, is the boundary of $M$  i.e., $\Sigma=\partial M$, it follows, by Stokes' theorem,
\begin{eqnarray}
 -\frac{1}{2} \int_{M} tr(R\wedge R) &=& -\frac{1}{2}\int_{\partial M=\Sigma} tr(\Gamma d\Gamma +\frac{2}{3}\Gamma \wedge\Gamma \wedge\Gamma)  \nonumber\\
&=& \frac{1}{2}\int_\Sigma {\tilde\epsilon}^{ikj} ( \Gamma^{m}_{in}\partial_{j}\Gamma^{n}_{km}+\frac{2}{3}\Gamma^{m}_{in}\Gamma^{n}_{js}\Gamma^{s}_{km} )d^{3}x  \nonumber\\
 & =& W_{CS}[\Gamma|_\Sigma];
 \end{eqnarray}
 wherein  $\Gamma^i_j =\Gamma^i_{jk}dx^k$ denotes the connection one-form on the spatial manifold $\Sigma$.
 However, $tr(R\wedge R)$ can only be locally exact in general; its associated Chern–Simons three-form can be related by a ``large gauge transformation'' with a nontrivial winding number.

\end{document}